\documentclass[useAMs,usenatbib,a4paper]{mn2e}
\usepackage{savesym}
\usepackage{graphicx}
\expandafter\let\csname equation*\endcsname\relax
  \expandafter\let\csname endequation*\endcsname\relax 
\usepackage{subfig}
\usepackage{amsmath}
\usepackage{verbatim}
\usepackage{array}
\usepackage{times}
\usepackage[total={17.8cm,24.0cm},centering]{geometry} 

\newcommand{\be}{\begin{equation}}
\newcommand{\ee}{\end{equation}}
\newcommand{\eea}{\end{eqnarray}}
\newcommand{\bea}{\begin{eqnarray}}

\newcommand{\m}{\mathrm}

\title[BL Lacs and the blazar sequence]{Synchrotron and inverse-Compton emission from blazar jets - IV. BL Lac type blazars and the physical basis for the blazar sequence}
\author[William J. Potter and Garret Cotter]{William J. Potter\thanks{E-mail:
will.potter@astro.ox.ac.uk (WJP)} and Garret Cotter
\\
Oxford Astrophysics. Denys Wilkinson Building, Keble Road, Oxford, OX1 3RH, United Kingdom}
\begin{document}

\date{}

\pagerange{\pageref{firstpage}--\pageref{lastpage}} \pubyear{2011}

\maketitle

\label{firstpage}

\begin{abstract}

In this paper we investigate the properties of a sample of six BL Lacs by fitting their spectra using our inhomogeneous jet model with an accelerating, magnetically dominated, parabolic base, which transitions to a slowly decelerating conical jet with a geometry based on observations of M87. Our model is able to fit very well to the simultaneous multiwavelength spectra of all the BL Lacs including radio observations. We find that the BL Lacs have lower jet powers and bulk Lorentz factors than the sample of Compton-dominant blazars investigated in Paper III, consistent with the blazar sequence. Excitingly, we find a correlation between the radius at which the jet first comes into equipartition and the jet power, in agreement with our prediction from Paper II. We interpret this result as one of two physical scenarios: a universal jet geometry which scales linearly with black hole mass or a dichotomy in Eddington accretion rates between flat spectrum radio quasars (FSRQs) and BL Lacs. If we assume that the jet geometry of all blazars scales linearly with black hole mass then we find a plausible range of masses ($\sim 10^{7}M_{\odot}-10^{10}M_{\odot}$). We find that the quiescent gamma-ray spectrum of Markarian 421 is best fitted by scattering of external CMB photons. We are unable to fit the spectrum using synchrotron self-Compton emission due to strong gamma-ray absorption via pair production even using a compact, rapidly decelerating, jet with a very large bulk Lorentz factor (50), as has been suggested recently. This is because the ratio of synchrotron to inverse-Compton emission requires a high density of synchrotron photons to scatter which makes the region opaque to TeV gamma-rays even with large bulk Lorentz factors. Finally, we fit to the SEDs of the four high power HSP BL Lacs recently found by Padovani et al. 2012. We find that their high peak frequency emission is caused by high maximum electron energies whilst the rest of their jet properties are typical of relatively high power BL Lacs and consistent with our predictions. 

\end{abstract}


\section{Introduction}

Blazars are thought to be supermassive black holes accreting matter and producing relativistic plasma jets oriented close to our line of sight. The observed spectra of blazars are dominated by the Doppler-boosted non-thermal synchrotron and inverse-Compton emission of electrons in the jet. The characteristic double-peaked spectrum is due to synchrotron emission which extends from radio to optical/X-ray energies and inverse-Compton emission which extends from X-ray to high energy $\gamma$-ray energies. Blazars are the most luminous and highly variable of active galactic nuclei (AGN) with radio emission that is observed to be nearly flat in flux due to synchrotron self-absorption.

The purpose of this series of papers is to develop a realistic, observationally motivated jet model which is able to reproduce the observed spectra of blazars across all wavelengths, so that we can learn about the physical properties of the population and the mechanisms involved in jet production. In the first two papers in the series (\cite{2012MNRAS.423..756P} and \cite{2013MNRAS.429.1189P}, hereafter referred to as Papers I and II) we developed a jet model with an accelerating parabolic base transitioning to a slowly decelerating conical jet with a geometry set by radio observations of M87 (\cite{2012ApJ...745L..28A}). We assume that the jet starts magnetically dominated and accelerates in the parabolic region until it reaches equipartition between particle and magnetic energy. Here the jet transitions from parabolic to conical at a distance $10^{5}$ Schwarzschild radii ($R_{s}$) and radius $2000R_{s}$ (the transition region) as observed in the jet of M87. We inject electrons (hereafter electron is used to refer to both electrons and positrons) at the transition region so that the plasma is in equipartition and we allow the jet to slowly decelerate in the conical region by converting bulk kinetic energy into particle acceleration, consistent with the in situ acceleration required to explain observations of optical synchrotron radiation at large distances along jets (\cite{1997A&A...325...57M} and \cite{2001A&A...373..447J}).

In our model we conserve relativistic energy-momentum and particle flux along the jet and treat emission processes thoroughly. We calculate the synchrotron and inverse-Compton emission by Lorentz transforming into the rest frame of the plasma and dividing the jet into small sections. We integrate the line of sight synchrotron opacity through the jet to each section and we use the full Klein-Nishina cross section to calculate the inverse-Compton emission from synchrotron, accretion disc, broad line region (BLR), dusty torus, narrow line region (NLR), CMB and starlight seed photons. We include radiative and adiabatic losses to the electron population travelling along the jet.  

In Papers II and III (\cite{2013MNRAS.431.1840P}) we investigated the 7 most Compton-dominant blazars in the \emph{Fermi} sample from \cite{2010ApJ...716...30A}. We found that all these blazars required high power, high bulk Lorentz factor jets with large radius transition regions and large black hole masses inferred from a jet geometry based on M87, scaled linearly with black hole mass. We found that our model fitted all these spectra very well across all wavelengths with sensible physical parameters. We found that the inverse-Compton emission of these blazars was best fitted by scattering of high-redshift CMB photons with a small contribution from NLR and starlight photons. This was the first time that scattering of CMB photons had been proposed as the reason for the Compton-dominance of high power blazars. We found that the optically thick to optically thin synchrotron turnover at relatively low frequencies was incompatible with a transition region (where the jet comes into equipartition and becomes synchrotron bright) within the BLR or dusty torus as is usually assumed in one-zone models (see for example \cite{2009MNRAS.399.2041G} and \cite{2010ApJ...721.1425A}).

We hypothesised that the blazar sequence (\cite{1998MNRAS.299..433F}) could be seen as a progression in jet power, black hole mass and bulk Lorentz factor as jet power increases. We set out a few simple arguments why this scenario is plausible and how it results in the decreasing peak frequency of synchrotron and inverse-Compton emission with increasing jet power due to the increasing size of the synchrotron bright transition region with increasing black hole mass. We further postulated that the most Compton-dominant blazars are observed at high-redshift since the energy density of the CMB is larger, resulting in higher Compton-dominance from scattering CMB photons at large distances along the jet. So a high power, high bulk Lorentz factor blazar at low redshift would be less Compton-dominant than the same blazar at high redshift. Finally, we predicted that BL Lac type blazars represent lower power blazars with low bulk Lorentz factors and low black hole masses. Smaller black hole masses result in smaller emission regions with higher magnetic field strengths so that the inverse-Compton emission of BL Lacs is due to SSC and both the synchrotron and inverse-Compton peak frequencies are larger than in the case of higher power blazars with larger emission regions and lower magnetic field strengths.

This is the first time that such a unified scenario has been suggested for blazars, based on a realistic, extended jet model consistent with observations, simulations and theory. In this paper we wish to test whether our model is capable of fitting to the multiwavelength observations of six BL Lac type blazars in order to test the validity of our model and the predictions we made in Paper III about physical properties of BL Lacs and their relation to the blazar population. In order to accurately model BL Lacs which can emit inverse-Compton radiation at very high energies we include the absorption of high energy gamma-rays along the jet due to photon-photon pair production in this work.

In this paper we will first briefly introduce our jet model from Paper II. We calculate the integrated optical depth along the jet for absorption of gamma rays due to pair production and show the results of fitting our model to six BL Lac type blazars. We then comment on the inferred physical parameters of the BL Lacs we find from fitting to their spectra and whether these results are consistent with our predictions for the blazar population from Paper III. We discuss the results and those from Paper III in the context of the blazar sequence and unification. We comment on the location of the high energy emission region in our jet model fit to the spectrum of Markarian 421 and investigate the constraints placed on the size of the region by the attenuation of gamma rays from pair production. Finally, we investigate the physical properties of four high power, high synchrotron peak frequency (HSP) BL Lacs by fitting our model to their SEDs and commenting on their relation to the blazar sequence.

\section{Jet Model}

We use as the basis of this investigation the jet model presented in Paper II. The jet model consists of an accelerating parabolic base transitioning to a slowly decelerating conical jet with a geometry set by the recent radio observations of the jet of M87 (\cite{2012ApJ...745L..28A}). We will briefly describe the principles of this model in the following section, however, for full details see Papers I, II and III.  

Our model is motivated by and consistent with observations, simulations and theory. The jet is assumed to start magnetically dominated at the base. The magnetic energy is converted into bulk kinetic energy in the plasma via a magnetic pressure gradient. The plasma is accelerated until it approaches equipartition between magnetic and particle energies where the jet approaches a terminal bulk Lorentz factor and transitions from parabolic to conical. In the conical section the jet slowly decelerates due to interactions with its environment (\cite{2005MNRAS.358..843H}) and this converts bulk kinetic energy into accelerating electrons via shocks. Deceleration is observed along jets and this is also consistent with evidence for in situ acceleration of electrons from observations of optical synchrotron emission (\cite{2001A&A...373..447J}). We assume that the plasma is isotropic and homogeneous in the rest frame of each section of plasma so can be described by a relativistic perfect fluid. We conserve energy-momentum and electron number via the conservation equations

\be
\nabla_{\mu}T^{\mu \nu}(x)=0, \qquad \nabla_{\mu}j_{e}^{\mu}(x)=0.
\ee

We calculate the synchrotron emission and opacity in each section by Lorentz transforming into the instantaneous rest frame. We integrate the synchrotron optical depth through the jet to each section to calculate the observed synchrotron emission. In the model we treat inverse-Compton scattering of synchrotron, CMB, starlight, accretion disc, BLR, dusty torus and NLR seed photons by Lorentz transforming into the plasma rest frame and using the full Klein-Nishina cross section. We evolve the electron population along the jet taking into account radiative losses from synchrotron and inverse-Compton emission and adiabatic losses due to expansion of the jet. We set the geometry of our model using the observations of the geometry of M87 scaled linearly with black hole mass.  

\subsection{Model parameters}

\begin{figure}
	\centering
		 \includegraphics[width=7 cm, clip=true, trim=4cm 1cm 8cm 4cm]{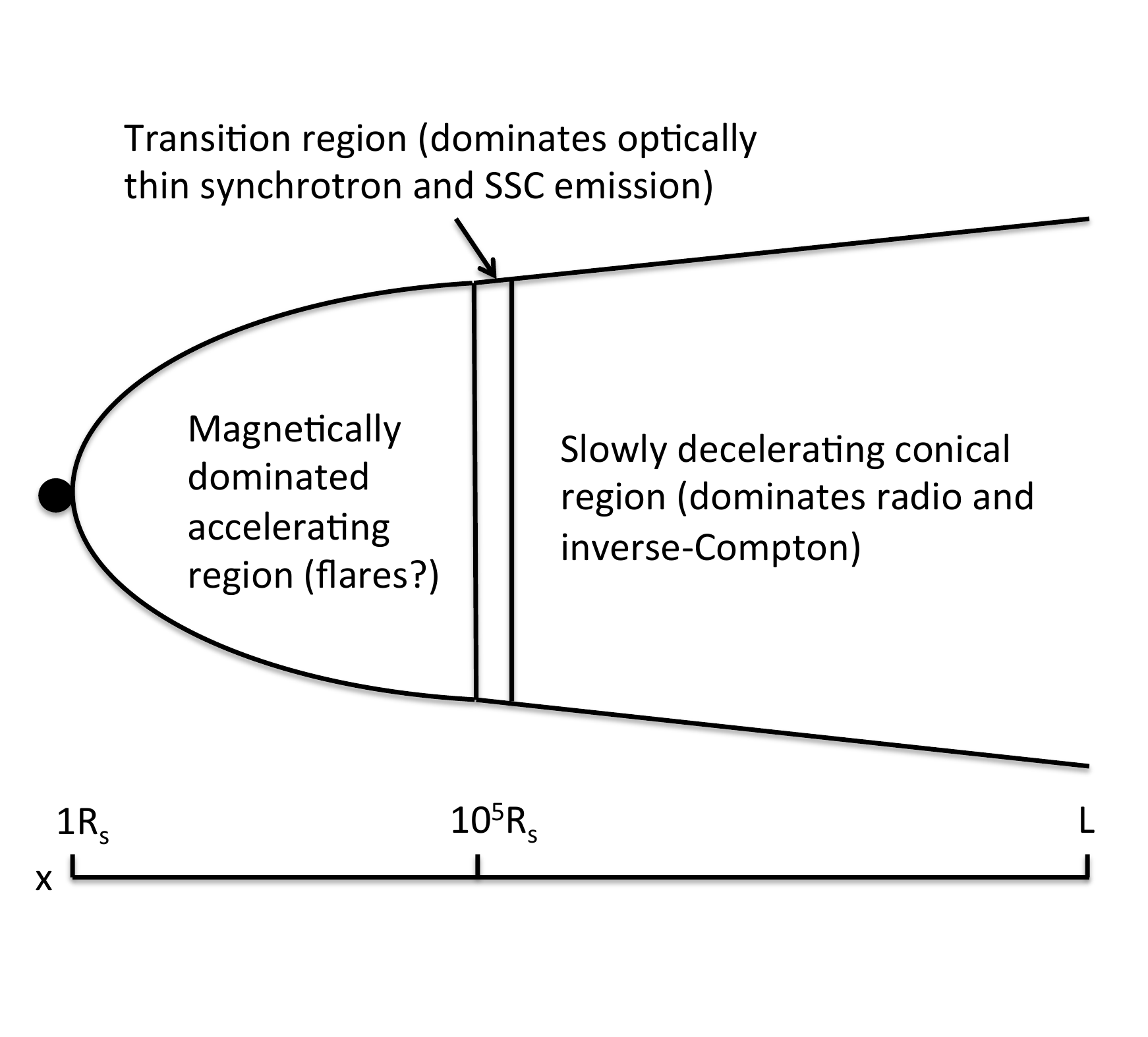} 
			
	\caption{A schematic diagram of our jet model (taken from Paper III). The jet starts magnetically dominated and accelerates in the parabolic region until the plasma reaches equipartition where the jet transitions to conical and slowly decelerates converting bulk kinetic energy into particle energy through shocks.}
	\label{fig:jet}
\end{figure}

We consider a jet with total lab frame power $W_{j}$, jet length $L$ and conical half-opening angle $\theta\rq{}_{\m{opening}}$ observed at an angle $\theta_{\m{observe}}$ to the jet axis. We assume that electrons accelerated along the jet have an initial distribution, $N_{e}(E_{e})\propto E_{e}^{-\alpha} \m{e}^{-E_{e}/E_{\m{max}}}$, with a minimum energy $E_{\m{min}}$. The base of the jet has an initial bulk Lorentz factor $\gamma_{0}$, which accelerates to $\gamma_{\m{max}}$ at the end of the parabolic section. The jet slowly decelerates from a bulk Lorentz factor of $\gamma_{\m{max}}$ to $\gamma_{\m{min}}$ by the end of the conical section at a lab frame distance L. The central black hole has a mass $M$. The accretion disc has a luminosity $L_{\m{acc}}$ and the outer distance of the NLR is given by $x_{\m{outer}}$. A schematic diagram of the jet is shown in Figure $\ref{fig:jet}$.

\section{Photon-photon pair production absorption}

In this paper we model the spectra of BL Lac type blazars. Unlike the powerful Compton-dominant blazars that we modelled in Paper III, BL Lac blazars have observed spectra which can extend up to TeV energies. For these very high energy photons we need to take into account the production of electron-positron pairs from collisions between these high energy photons and ambient photons. When such a collision occurs the two photons are converted into an electron pair where each electron has approximately half the energy of the high energy photon and travels in approximately the same direction as the high energy photon. For two photons with energies $E_{1}$ (high energy photon) and $E_{2}$ (ambient photon) and velocities forming an angle $\theta$ to each other, the threshold energy of the ambient photon $E_{\m{thresh}}$ above which pair production can occur is given by

\be
E_{\m{thresh}}=\frac{2 m_{e}^{2}c^{4}}{E_{1}(1-\cos \theta)}.
\ee

The total cross section for pair production is given by (\cite{1967PhRv..155.1404G})

\be
\sigma_{\gamma \, \gamma}=\frac{1}{2}\pi r_{0}^{2}(1-\beta_{e}^{2})\left[(3-\beta_{e}^{4})\ln \left(\frac{1+\beta_{e}}{1-\beta_{e}}\right)-2\beta(2-\beta_{e}^{2})\right].
\ee

\be
\beta_{e}=\left(1-\frac{2m_{e}^{2}c^{4}}{E_{1}E_{2}(1-\cos \theta)}\right)^{1/2}, \qquad r_{0}=\frac{e^{2}}{4\pi \epsilon_{0}m_{e}c^{2}},
\ee

where $\beta_{e}$ is the electron velocity in the centre of momentum frame divided by c and $r_{0}$ is the classical electron radius. We wish to calculate the fraction of high energy photons absorbed through pair production as they travel along the jet. We calculate the pair production optical depth of each section of the jet in the rest frame of the section. The total optical depth for a photon of energy $E_{1}$ emitted at an angle $\theta_{\m{observe}}$ to the jet axis in the lab frame from a section at lab frame distance $x\rq{}$ along the jet due to synchrotron photons is

\bea
&\tau_{\gamma \gamma 1}(E_{1},x')&= \sum_{x=x'}^{x=L} \sum_{E_{2}=E_{\m{thresh}}}^{E_{2}=\infty} \sum_{\theta=0}^{\theta=\pi} \m{d}r(x) (1-\cos \theta) ...
\nonumber \\
&&\times  n_{\m{synch}}(E_{2},x)\frac{\sin \theta}{2} \sigma_{\gamma \gamma}(E_{1}\rq{}(x),E_{2},\theta) \m{d}E_{2} \m{d} \theta,\nonumber
\eea 
\be
\m{d}r(x)=\gamma_{\m{bulk}}(x)^{2}\left(\frac{1}{\cos \theta_{\m{observe}}}-\beta(x)\right) \m{d}x',
\ee
\be
E_{1}\rq{}(x)=\frac{\delta_{\m{Dopp}}(x\rq{})}{\delta_{\m{Dopp}}(x)}E_{1}, \,\,\, \delta_{\m{Dopp}}(x)=\frac{1}{\gamma_{\m{bulk}}(x)(1-\beta(x)\cos \theta_{\m{obs}})}
\ee

where $n_{\m{synch}}(E_{2},x)$ is the number density of synchrotron photons in the plasma rest frame and $\m{d}r(x)$ is the distance travelled by the photon through a section of rest frame width $\m{d}x'$ in the rest frame of the plasma (see Paper I). We have taken into account the relative Doppler-boosting of the initial photon of energy $E_{1}$ into sections with different bulk Lorentz factors by defining the rest frame photon energy $E_{1}\rq{}(x)$ to be the photon energy in the relevant section of the jet. We have assumed that the synchrotron photons are isotropically distributed in the rest frame of the plasma and we have integrated the total optical depth through the jet to the section at a lab frame distance $x$.

The total optical depth for pair production due to CMB, BLR, dusty torus, NLR and starlight photons is

\bea
\tau_{\gamma \gamma 2}(E_{1},x')= \sum_{x=x'}^{x=L} \sum_{E_{2}=E_{\m{thresh}}}^{E_{2}=\infty} \m{d}r(x) (1-\cos \theta) n_{\m{EC}}(E_{2},x)... \nonumber \\
\times \sigma_{\gamma \gamma}(E_{1}',E_{2},\theta) \m{d}E_{2},
\eea

\be
\theta=\pi-\cos^{-1}\left(\frac{\cos \theta_{\m{obs}}-\beta(x)}{1-\beta(x)\cos \theta_{\m{obs}}}\right),
\ee

\bea
n_{\m{EC}}(E_{2},x)=n_{\m{CMB}}(E_{2},x)+n_{\m{BLR}}(E_{2},x)+n_{\m{star}}(E_{2},x)... \nonumber \\
+n_{\m{dusty}\, \m{torus}}(E_{2},x)+n_{\m{NLR}}(E_{2},x),
\eea

where we have Lorentz transformed the CMB, BLR, dusty torus, NLR and starlight photons into the plasma rest frame and assumed that the photons are strongly Doppler-boosted so that the majority of photons travel approximately parallel to the jet axis (see Paper II). Finally, the total optical depth due to accretion disc photons is given approximately by

\bea
\tau_{\gamma \gamma 2}(E_{1},x')= \sum_{x=x'}^{x=L} \sum_{E_{2}=E_{\m{thresh}}}^{E_{2}=\infty} \m{d}r(x) (1-\cos \theta) n_{\m{acc}}(E_{2},x) \nonumber \\
\times \sigma_{\gamma \gamma}(E_{1}',E_{2},\theta) \m{d}E_{2},
\eea

\be
\theta=\cos^{-1}\left(\frac{\cos \theta_{\m{obs}}-\beta_{x}}{1-\beta(x)\cos \theta_{\m{obs}}}\right),
\ee

where we have Lorentz transformed the number density of accretion disc photons into the plasma rest frame as calculated in Paper II. The total optical depth due to pair production from all these photon fields is given by

\be
\tau_{\gamma \gamma \m{tot}}(E_{1},x\rq{})=\tau_{\gamma \gamma 1}(E_{1},x\rq{})+\tau_{\gamma \gamma 2}(E_{1},x\rq{})+\tau_{\gamma \gamma 3}(E_{1},x\rq{}).
\ee

So the fraction of high energy photons with energy $E_{1}$ emitted from a section of the jet at lab frame distance $x$ which are observed is

\be
\m{fraction}(E_{1},x)=\m{e}^{-\tau_{\gamma \gamma \m{tot}}(E_{1},x\rq{})},
\ee

In the current work we neglect the subsequent emission of pair produced electrons, this is justified since in synchrotron-dominated high peak frequency BL Lacs (HBLs) only a small fraction of the total particle energy is emitted via high energy photons. The conversion of high energy photons into electron-positron pairs will reduce the amount of high energy emission observed from HBLs. It is important to consider pair production when modelling HBLs since the observation of TeV photons can be used to constrain the size of the high energy emitting region. This is because compact emission regions located close to the base of the jet will have a larger number density of both synchrotron and ambient photons than larger regions at larger distances emitting the same power and so compact regions may be opaque to TeV photons. Constraints on the location, size and bulk Lorentz factors of high energy emission regions in one-zone models have been investigated by many authors including \cite{2008MNRAS.384L..19B}, \cite{2009ApJ...703.1168B} and \cite{2012ApJ...755..147D}. This is the first investigation which calculates the effect of pair production in an extended jet model by Lorentz transforming ambient photon fields into the plasma rest frame and integrating the pair production optical depth along the jet. 

\section{External photon fields for BL Lacs}

It is generally observed that BL Lac spectra are dominated by a featureless synchrotron continuum at optical wavelengths. It is thought that this is due to a weak or absent BLR and dusty torus, and synchrotron emission which dominates the accretion disc spectrum (\cite{1995ApJ...452L...5V}). In contrast FSRQ spectra have emission line features with broad emission lines (\cite{2011MNRAS.414.2674G}). The inverse-Compton emission of BL Lacs is normally attributed to SSC emission whilst the inverse-Compton emission of FSRQs is thought to be primarily due to scattering external photons from the BLR and dusty torus (\cite{2012ApJ...752L...4M}). In Paper III we found that the inverse-Compton emission of a sample of Compton-dominant blazars was incompatible with SSC emission but was well fitted by scattering of external photons (CMB and NLR photons). In this paper we shall initially assume that the inverse-Compton emission of our BL Lacs is due to SSC emission and neglect BLR, accretion disc, dusty torus and NLR photons before attempting to fit the spectra using external photons. It is interesting to see whether these BL Lac spectra can be fitted with SSC emission or whether they require external photon fields to fit the spectra. 

The jets of BL Lacs are thought to correspond to FRI jets and so we shall attempt to fit the BL Lac spectra with shorter jets than in Paper III where we were fitting to powerful Compton-dominant sources.
 
\section{Results}

In this section we show the results of fitting our model to the multi-wavelength observations of six BL Lac type blazars (J$0222.6+4302$, J$0855.4+2009$, J$1015.2+4927$, J$1104.5+3811$, J$1751.5+0935$ and J$2158.8 - 3014$) selected by the availability of redshifts and good quality simultaneous multiwavelength spectra across radio, optical, X-ray and $\gamma$-rays from \cite{2010ApJ...716...30A}. The spectrum for Mkn421 from \cite{2010ApJ...716...30A} contains multiple sets of data and so we have instead used the simultaneous multiwavelength data from the \emph{Fermi} campaign on Mkn421 (\cite{2011ApJ...736..131A}). 

\begin{figure*}
	\centering
		\subfloat[J0222]{ \includegraphics[width=8 cm, clip=true, trim=1cm 1cm 0cm 1cm]{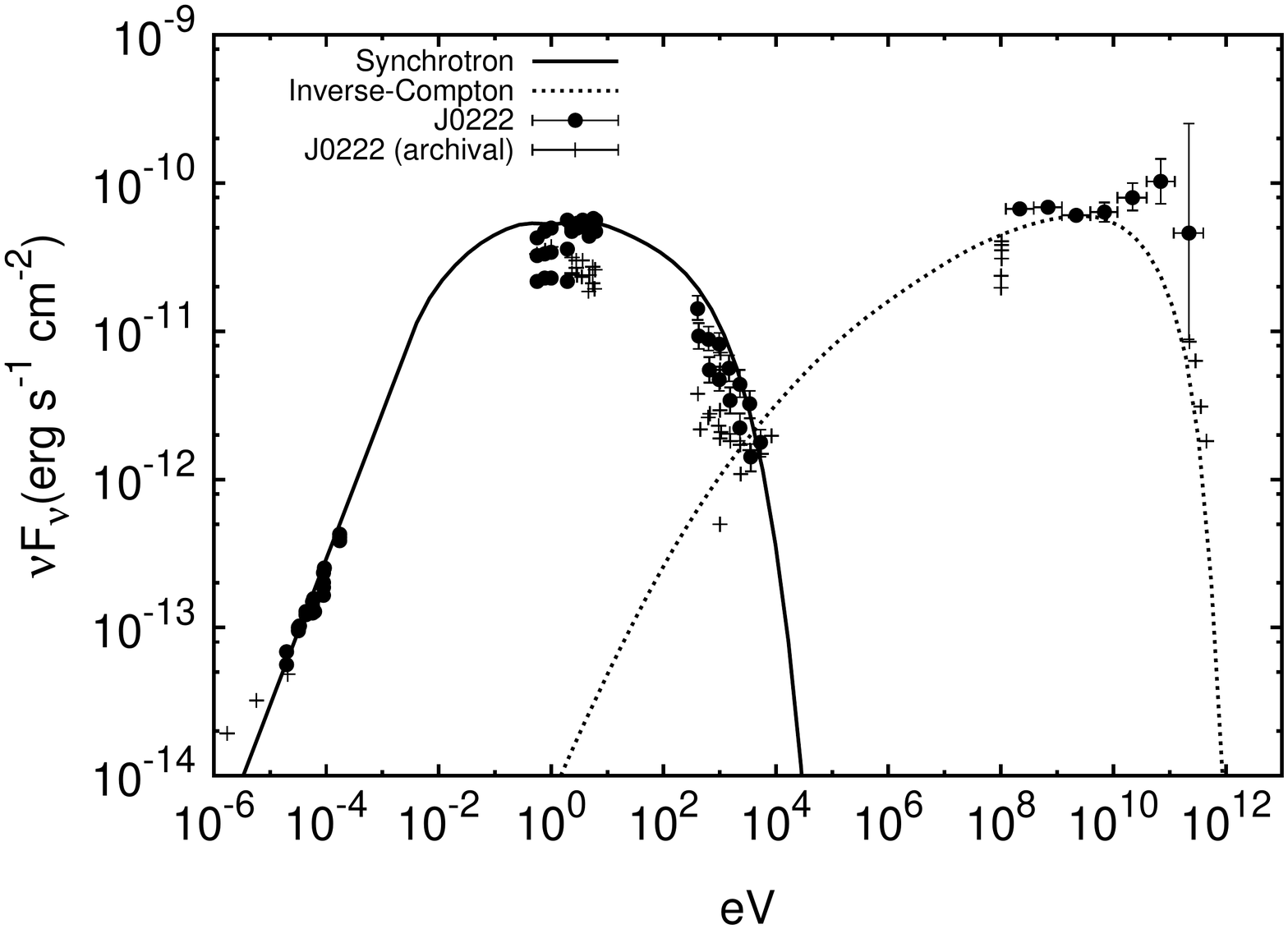} }
		\qquad
		\subfloat[J0855]{ \includegraphics[width=8cm, clip=true, trim=1cm 1cm 0cm 1cm]{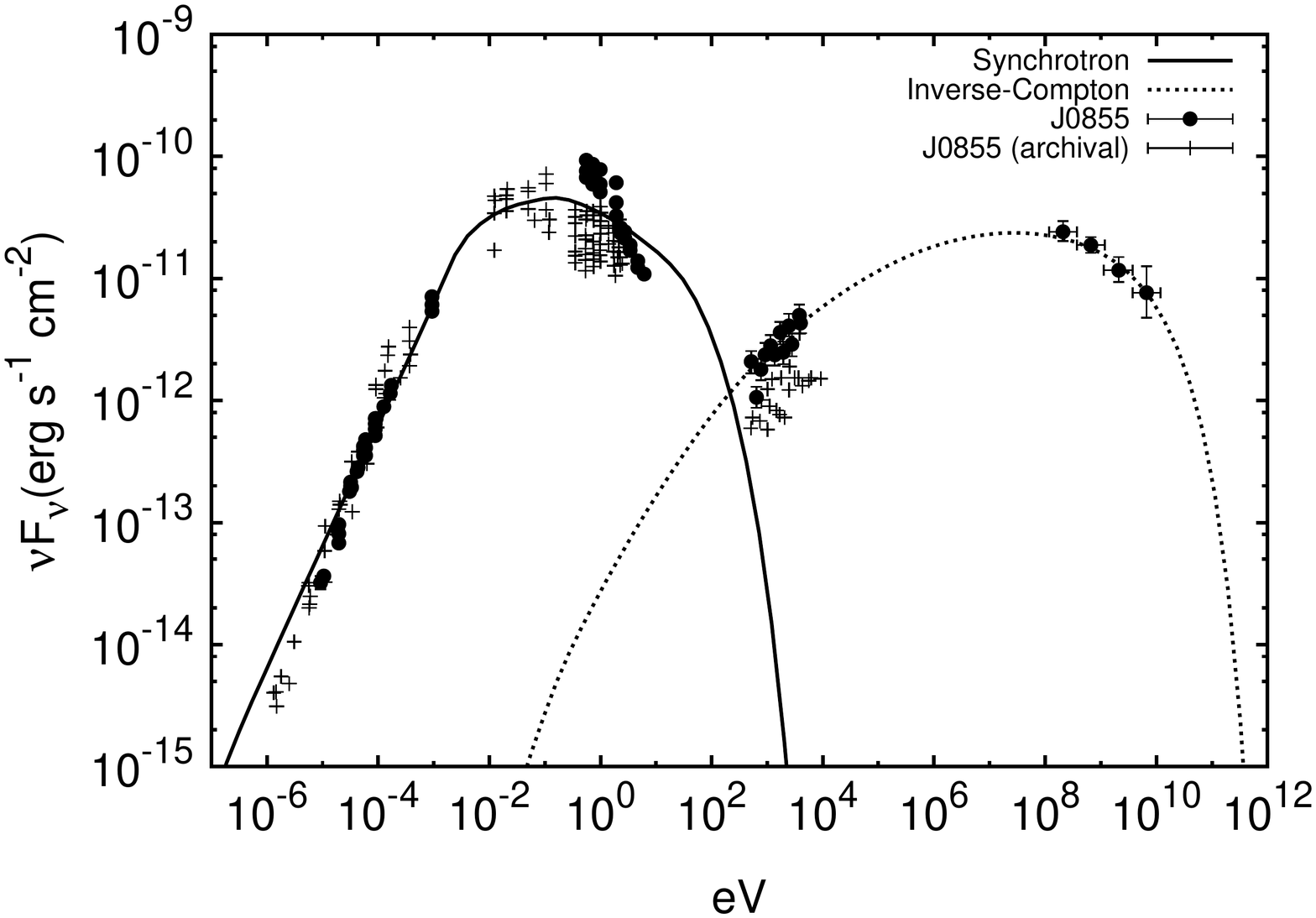} }
		\\
		\subfloat[J1015]{ \includegraphics[width=8cm, clip=true, trim=1cm 1cm 0cm 1cm]{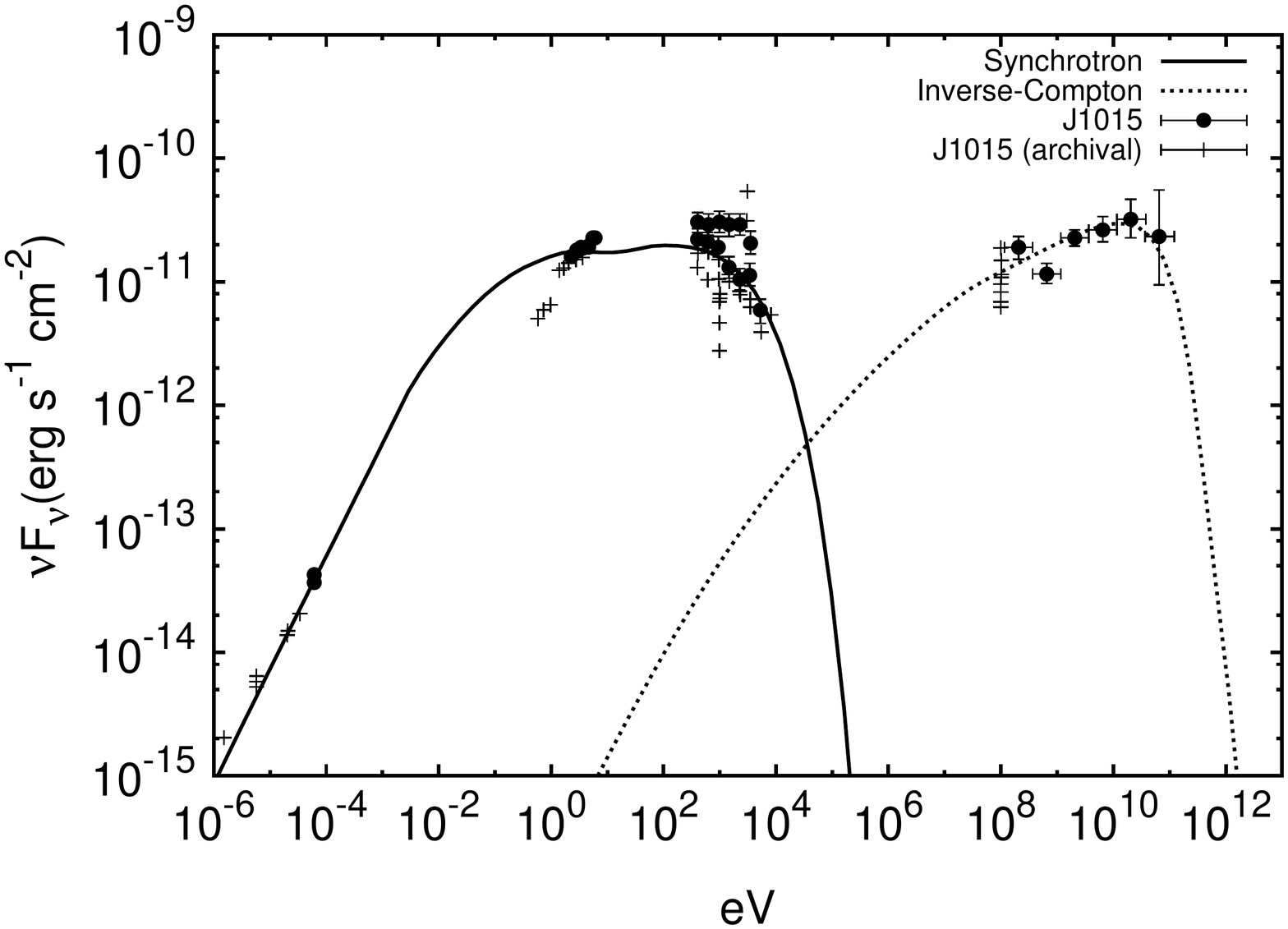} }
		\qquad
		\subfloat[Mkn421]{ \includegraphics[width=8 cm, clip=true, trim=1cm 1cm 0cm 1cm]{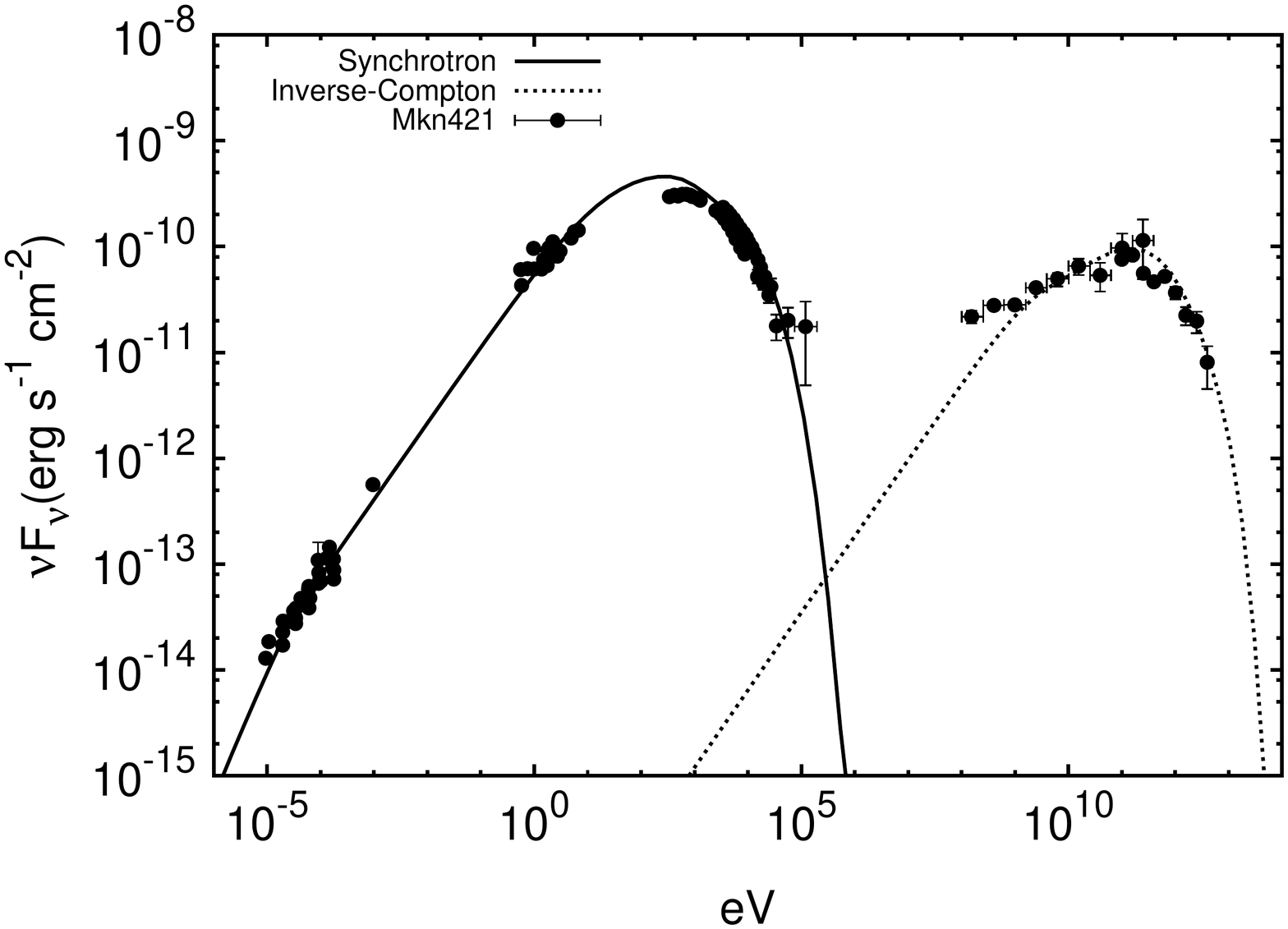} } 
		\\
		\subfloat[J1751]{ \includegraphics[width=8 cm, clip=true, trim=1cm 1cm 0cm 1cm]{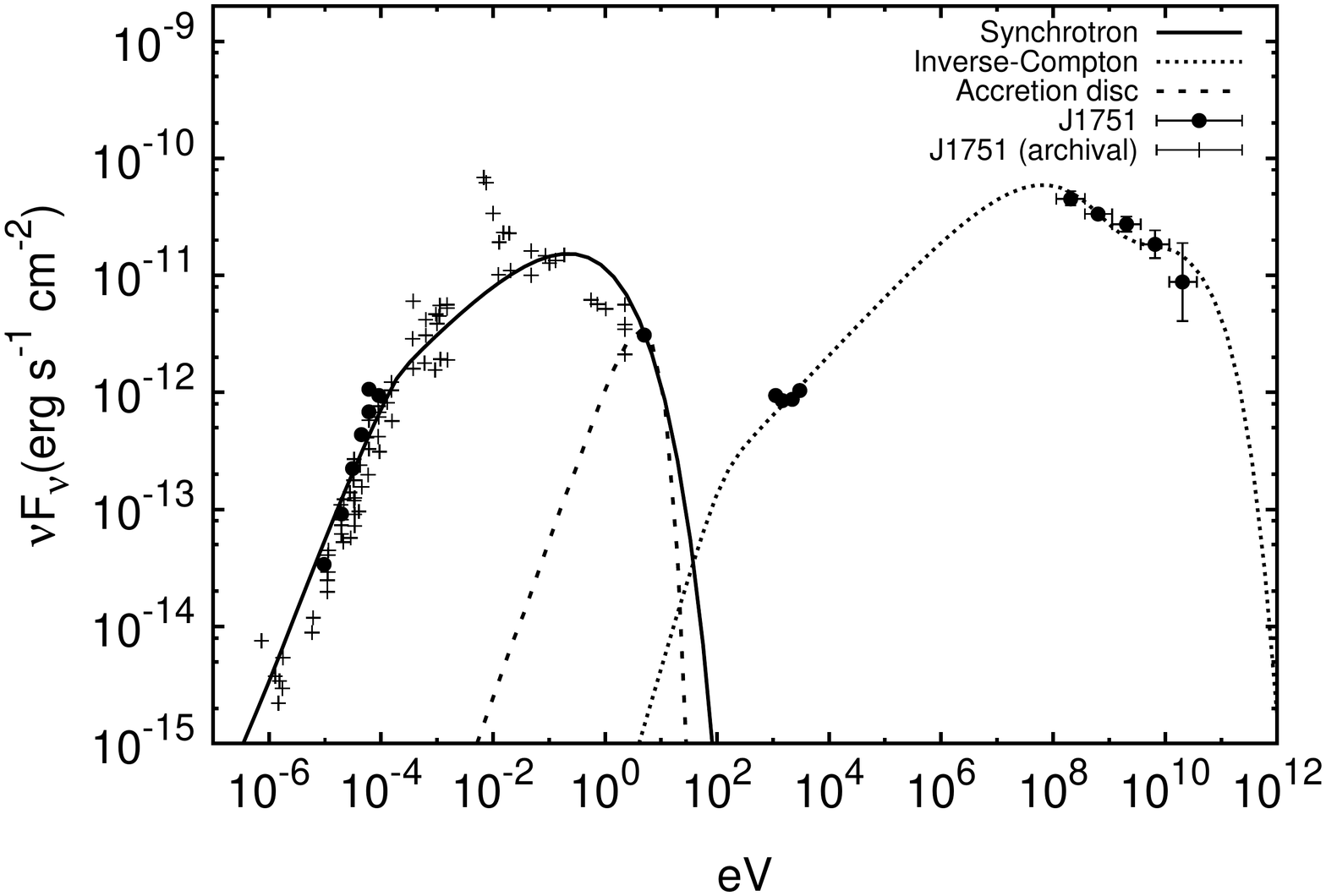} }
		\qquad
		\subfloat[J2158]{ \includegraphics[width=8 cm, clip=true, trim=1cm 1cm 0cm 1cm]{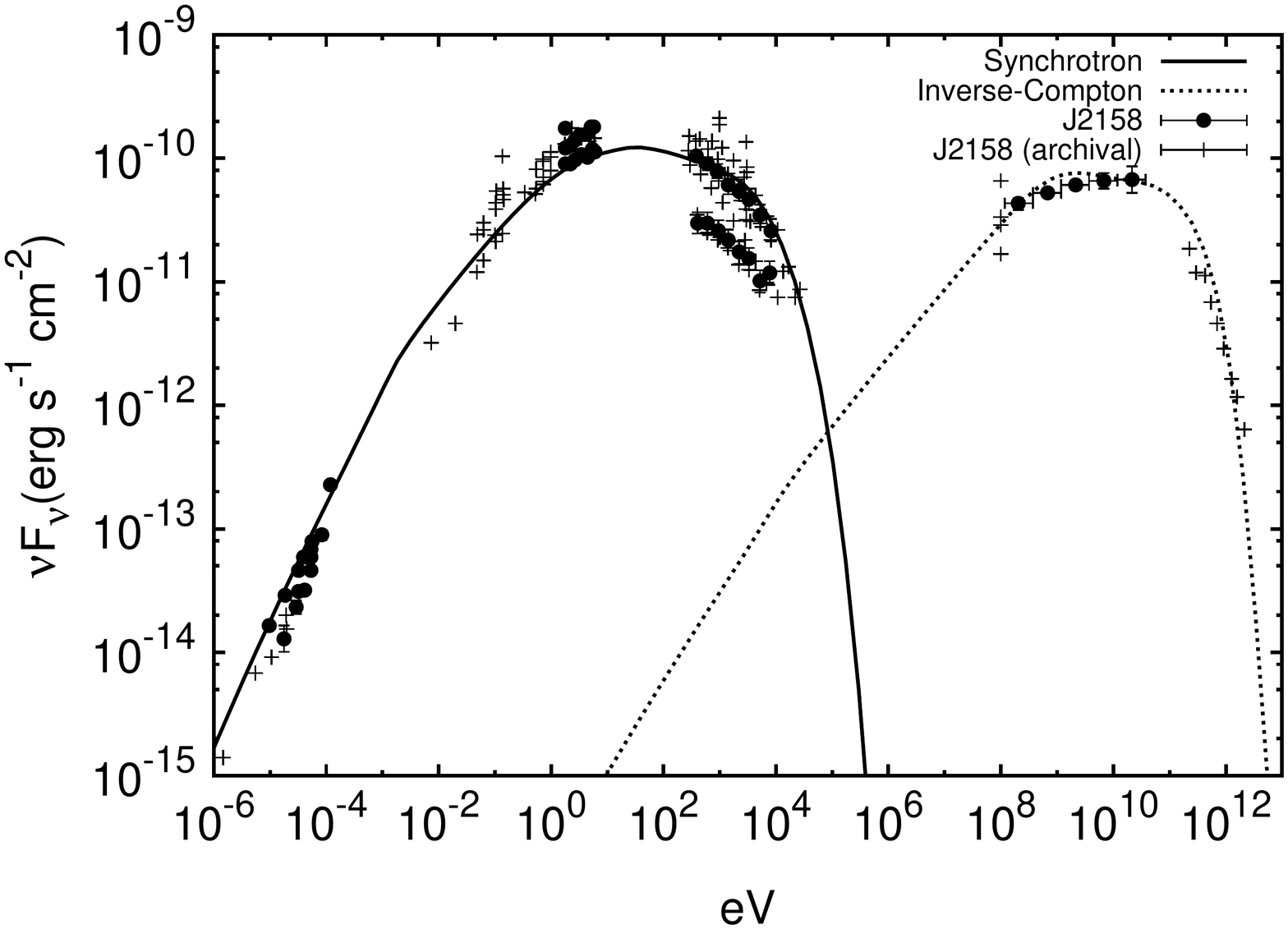} }
			
	\caption{The results of fitting our model to the SEDs of J0222, J0855, J1015, Mkn421, J1751 and J2158. The model fits the observations well for all objects across all wavelengths. We find that the inverse-Compton emission of all the blazars, except J1751 and Mkn421, are well fitted by SSC emission. We find that the low peak frequencies of J1751 imply that the blazar is a more powerful Compton-dominant blazar at low redshift. }
	\label{fig:1}
\end{figure*}

\begin{figure*}
	\centering
		\subfloat[J1751]{ \includegraphics[width=14 cm, clip=true, trim=1cm 1cm 0cm 1cm]{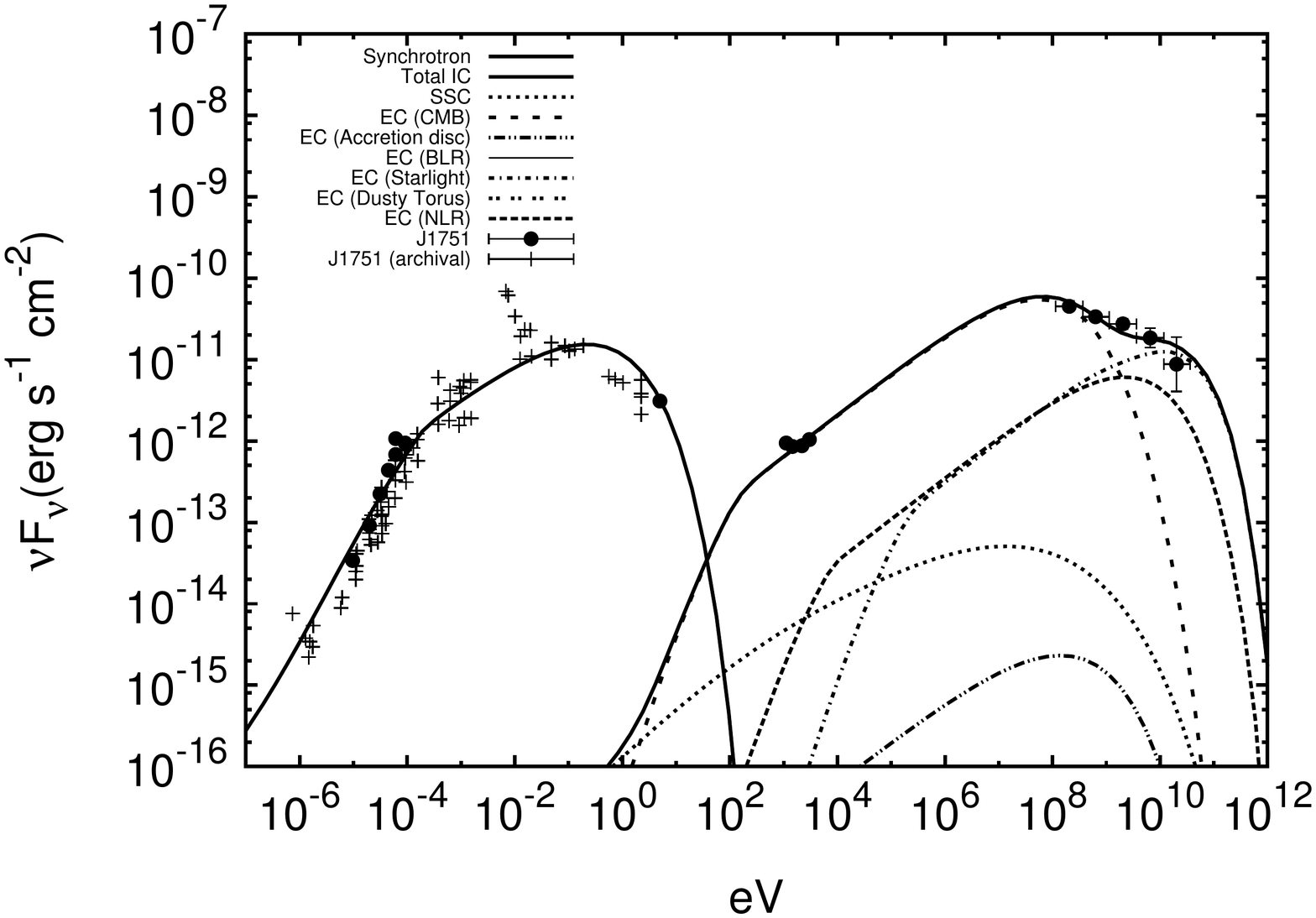} }
			
	\caption{The different components of the inverse-Compton emission for the fit to J1751. The inverse-Compton emission is primarily due to scattering of CMB seed photons with a small contribution of starlight required to fit to the highest energy data-point. For these Compton-dominant blazars we find that the SSC emission is sub-dominant due to the comparatively low magnetic field strength at the transition region of the jet.}
	\label{fig:2}
\end{figure*}

\begin{table*}
\centering
\begin{tabular}{| c | c | c | c | c | c | c | c |}
\hline
Parameter & J0222 & J0855 & J1015 & Mkn421 & J1751 & J2158 \\ \hline 
$W_{j}$ & $1.0 \times 10^{38}\rm{W}$ & $1.6\times 10^{38}\rm{W}$ & $2.4 \times 10^{37}\rm{W}$ & $6.0 \times 10^{36}\rm{W}$ & $2.1 \times 10^{38}\rm{W}$ & $1.6 \times 10^{37}\rm{W}$  \\ \hline
L & $5 \times 10^{20}\rm{m}$ & $5 \times 10^{20}\rm{m}$ & $1 \times 10^{21}\rm{m}$ & $3 \times 10^{20}\rm{m}$ & $2 \times 10^{21}\rm{m}$ & $1 \times 10^{21}\rm{m}$ \\ \hline
$E_{\m{min}}$ & 5.11 \rm{MeV} & 5.11 \rm{MeV} & 5.11 \rm{MeV} & 5.11 \rm{MeV} & 5.11 \rm{MeV} & 5.11 \rm{MeV} \\ \hline
$E_{\m{max}}$ & 16.2 \rm{GeV} & 5.11 \rm{GeV} & 32.2 \rm{GeV} & 338 \rm{GeV} & 4.06 \rm{GeV} & 51.1 \rm{GeV}  \\ \hline
$\alpha$ & 1.6  & 1.63  & 1.5 & 1.55 & 1.98 & 1.7  \\ \hline
$\theta'_{\m{opening}}$ & $3^{o}$ & $5^{o}$ & $5^{o}$ & $5^{o}$ & $3^{o}$ & $5^{o}$ \\ \hline
$\theta_{\m{observe}}$ & $1.0^{o}$ & $1.0^{o}$ & $3.5^{o}$ & $3.0^{o}$ & $1.0^{o}$ & $2.0^{o}$ \\ \hline
$\gamma_{\m{0}}$ & 4 & 4 & 4 & 4 & 4 & 4 \\ \hline
$\gamma_{\m{max}}$ & 12 & 8 & 10 & 12 & 34 & 15 \\ \hline
$\gamma_{\m{min}}$ & 8 & 6 & 3 & 9 & 18 & 7 \\ \hline
$R_{T}$ & $2.5 \times 10^{14}\m{m}$ & $6.8 \times 10^{14} \m{m}$ & $9.8 \times 10^{13} \m{m}$ & $3.7 \times 10^{15} \m{m}$ & $1.1 \times 10^{16} \m{m}$ & $1.7 \times 10^{14} \m{m}$ \\ \hline
$M$ & $4.2 \times 10^{7}M_{\odot}$ & $1.2 \times 10^{8}M_{\odot}$ & $1.7 \times 10^{7}M_{\odot}$ & $6.3 \times 10^{8}M_{\odot}$ & $1.8 \times 10^{9}M_{\odot}$ & $2.9 \times 10^{7}M_{\odot}$ \\ \hline
$L_{\m{acc}}$ & $-$ & $-$ & $-$ & $-$ & $3.3 \times10^{37}\rm{W}$ & $-$ \\ \hline
$x_{outer}$ & $-$ & $-$ & $-$ & $-$ & $1\m{kpc}$ & $-$ \\ \hline
\end{tabular}
\caption{The values of the physical parameters used in the model fits shown in Figure $\ref{fig:1}$.}
\label{tab1}
\end{table*}

In our model we have initially assumed that the size of the jet scales linearly with black hole mass with a transition region that occurs at a distance $10^{5}R_{s}$ with a radius $2000R_{s}$ set from radio observations of M87. In fitting our model to a spectrum we find a radius for the transition region, $R_{T}$, which gives us an inferred black hole mass, $M$, for each fit. We discuss the assumption of a jet geometry which scales linearly with black hole mass and the implications of alternative scenarios in sections 5.2-5.4.

We show the results of fitting our model to the blazars in Figure $\ref{fig:1}$ and Table \ref{tab1}. The model fits very well to all the spectra across all wavelengths. We find that the inverse-Compton emission of the majority of blazars (except for J1751 and Mkn421) is due to SSC emission as is commonly found for BL Lac type objects (\cite{2010arXiv1006.5048B}). We find that these blazars (excluding J1751) are relatively low power, low bulk Lorentz factor objects with smaller radius transition regions (where the jets first reach equipartition) compared to the Compton-dominant blazars we fitted in Paper III. 

This is a significant result since it shows that our model is able to successfully fit to a variety of blazar spectra from different sections of the blazar population, including radio observations. It is exciting to find that the physical parameters of our fits are consistent with our predictions from Paper III and also with the blazar sequence (\cite{1998MNRAS.299..433F}). We find in agreement with the blazar sequence and previous investigations that BL Lac type objects are relatively low power blazars in which the inverse-Compton emission is predominantly due to SSC emission. In Paper III we found that Compton-dominant blazars were high power blazars in which the inverse-Compton emission was predominantly due to scattering of the CMB.  

We find that the blazar J1751 has a larger power, bulk Lorentz factor and transition region radius than the other BL Lac blazars. The physical parameters of the fit are more similar to the Compton-dominant blazars found in Paper III. We find that in order to reproduce the low frequency synchrotron peak the fit requires a large transition region (and a large inferred black hole mass if the transition region occurs at $10^{5}R_{s}$ as in M87). We are unable to fit the inverse-Compton emission with SSC emission due to the low magnetic field strength required to produce the low peak synchrotron frequency. We find that BLR and dusty torus photons are unable to reproduce the observed high energy emission and archival synchrotron data simultaneously. This is because the archival observations show the synchrotron emission becomes optically thin at the relatively low energy $\sim10^{-4}$eV, requiring a low magnetic field strength and large transition region (see Paper III Equation 6). This large transition region is located at a distance $18$pc from the black hole in our fit, placing it outside the BLR and dusty torus. Somewhat surprisingly, we find that the inverse-Compton emission of J1751 is well fitted by scattering of CMB photons even at the comparatively low redshift $z=0.322$, if the jet has a large bulk Lorentz factor (see Figure \ref{fig:2}). These characteristics indicate that the blazar has similar physical properties to the Compton-dominant blazars from Paper III but is located at low redshift where the CMB has a lower energy density.

In Paper III we found that the inverse-Compton emission of Compton-dominant blazars was best fitted by scattering of Doppler-boosted high redshift CMB photons. We predicted that if the inverse-Compton emission of high power Compton-dominant blazars was due to scattering of CMB photons, then at low redshift we should see blazars with the same physical parameters but with less Compton-dominance due to the redshifting of the energy density of the CMB ($\rho_{\m{CMB}}\propto (1+z)^{4}$). J1751 appears to be a blazar matching the description of a high power, high bulk Lorentz factor jet with a large transition region but a lower Compton-dominance due to its low redshift. We find this evidence for the inverse-Compton scattering of CMB photons intriguing and we intend to investigate whether other high power, high bulk Lorentz factor jets at low redshift are compatible with their inverse-Compton emission being mainly produced by scattering of CMB seed photons.

Whilst we might expect the jet power, bulk Lorentz factor and radius of the transition region to be related, it is not clear how we expect these parameters to impact the mechanism of particle acceleration in jets. For the high power Compton-dominant sources investigated in Paper III we found that the electron distribution power-law index was considerably lower than the value $\alpha=2$ predicted by diffusive parallel shock acceleration (\cite{1978MNRAS.182..147B}). We find that all of these blazars require power-law indices less than or equal to 2. Spectral indices less than 2 are predicted in the case of oblique shock acceleration (\cite{2011MNRAS.418.1208B} and \cite{2012ApJ...745...63S}) and magnetic reconnection (\cite{2007ApJ...670..702Z}). We find that the maximum electron energy tends to be lower in higher power objects and increases in lower power objects, however, we do not find an obvious physical reason for this. This trend is usually explained by stronger radiative cooling of accelerating electrons in higher power FSRQs with luminous BLRs (\cite{2009MNRAS.399.2041G}). In our model, however, we find that the region responsible for the high energy emission (the transition region) is relatively large for high power blazars and is located outside the BLR, so electrons are not strongly radiatively cooled. We find that for all the blazars (except Mkn421) the maximum electron energy ($E_{\m{max}}$) is between $4-55$GeV while Mkn421 requires a much larger maximum electron energy, $338$GeV, considerably higher than the other blazars. 

We find that the radio spectra of the blazars (excluding J1751) are close to flat in flux and so are better fitted with a ballisitic equipartition jet rather than an adiabatic equipartition jet or a ballistic jet starting in equipartition at the transition region and becoming more particle dominated further along the jet ($A_{\m{equi}}=2x_{T}/(x+x_{T})$ for $x_{T} \geq x \geq 100x_{T}$). In Paper III we found that the multiwavelength spectra were best fitted by a ballistic conical jet starting in equipartition at the transition region and becoming more particle dominated further along the jet. We wish to investigate whether, for a larger sample, BL Lacs are best fitted by jets with ballistic equipartition conical sections whilst FSRQs favour ballistic jets which become particle dominated along the conical section. This would imply an important, fundamental difference between BL Lac and FSRQ jets.

\subsection{Understanding the blazar sequence}

In Paper III we hypothesised that the blazar sequence could be explained if the radius of the transition region and bulk Lorentz factor of the jet were to increase with the jet power. In this paper and Paper III we have shown that FSRQs are well fitted by high power jets with large radius transition regions and large bulk Lorentz factors whilst BL Lacs are well fitted by lower power jets with smaller radius transition regions and lower bulk Lorentz factors. 

We have shown that our model is able to fit well to a sample of BL Lacs using a jet model which stops accelerating and comes into equipartition at smaller distances with a smaller jet radius than the sample of FSRQs fitted in Paper III. This is an important result which confirms the prediction of Paper III. The link between jet power and the jet radius at the transition region not only allows us to explain the physics behind the blazar sequence but also to learn how the jet dynamics are affected by the properties of the jet. Depending on the relationship between black hole mass and jet power we find two interpretations of our results: either a universal jet geometry which scales linearly with black hole mass or a dichotomy in accretion rate between BL Lacs and FSRQs.

\subsection{Scenario I - A universal jet geometry}

Our jet model is based on observations of the jet in M87 scaled linearly with black hole mass. We have shown that this model is able to fit well to both BL Lacs and FSRQs, if FSRQs have large black hole masses $10^{9}-10^{10}M_{\odot}$ with large bulk Lorentz factors and BL Lacs have smaller black hole masses $10^{7}-10^{9}M_{\odot}$ with smaller bulk Lorentz factors. This range of black hole masses coincides with the expected range of black hole masses for AGN (\cite{2008ApJ...674L...1V}). In the case of a few FSRQs with accretion disc spectra visible above the jet emission, it is possible to find an independent estimate of the black hole mass. In these cases we find that this black hole mass agrees with the black hole mass inferred from the jet model in which the transition region occurs at $10^{5}R_{s}$. Unfortunately, in BL Lacs the jet emission tends to overwhelm the accretion disc spectra and so it is difficult to estimate the black hole mass via this method. This scenario of a universal jet geometry would seem logical if blazars all accrete at approximately the same Eddington fraction, so that the accretion power depends linearly on the black hole mass. This scenario would predict a linear relation between the jet power and the jet radius at the transition region. 

Black hole mass estimates for blazars are mostly indirect and suffer from large uncertainties in their calibration (\cite{2002ApJ...579..530W}, \cite{2010MNRAS.409..591F}, \cite{2011ApJS..194...45S} and \cite{2012ApJ...748...49S}). However, systematic studies of FSRQ and BL Lac spectra using black hole mass estimates based on emission line widths and virial estimates have found that the black hole masses of FSRQs and BL Lacs do not seem to differ substantially, with BL Lacs, in fact, having larger black hole masses on average \cite{2013ApJ...764..135S}. If we believe these measurements this leads us to a second interpretation of our results.

\subsection{Scenario II - A dichotomy in accretion rate}

If there is no systematic difference between the black hole masses in FSRQs and BL Lacs then our results indicate that the transition region occurs at a distance and radius which depends mainly on the jet power (or accretion rate) and is not strongly dependent on the black hole mass. We have found that our sample of BL Lacs have lower jet powers and smaller radius transition regions than FSRQs. If the black hole masses of FSRQs and BL Lacs are similar this means that the jet power in BL Lacs is a smaller fraction of the Eddington luminosity than in FSRQs, as has been suggested by \cite{2008MNRAS.387.1669G}. This would support the idea of a dichotomy in the accretion modes of FSRQs and BL Lacs. It has been previously suggested that FSRQs have large Eddington accretion rates with radiatively efficient thin discs while BL Lacs have much lower Eddington accretion rates and radiatively inefficient discs (see for example \cite{2009MNRAS.396L.105G} and \cite{2011ApJ...740...98M}). We can make a rough estimate of the Eddington luminosity by using a fiducial black hole mass for BL Lacs of $5 \times 10^{8} M_{\odot}$ taken from \cite{2013ApJ...764..135S}. Using this mass we estimate a range of luminosities $0.001L_{\m{Edd}}-0.03L_{\m{Edd}}$ for the BL Lacs fitted in this paper. Taking the same fiducial mass for the FSRQs fitted in Paper III we estimate a range of luminosities $0.2L_{\m{Edd}}-0.9L_{\m{Edd}}$.

Our results would link this dichotomy in accretion mode to the properties of the jet produced in the two modes. FSRQs accreting at high rates produce jets which accelerate over larger distances reaching larger bulk Lorentz factors and radii before coming into equipartition at the transition region than the jets of BL Lacs. This would also be consistent with AGN unification in which BL Lacs correspond to FRI type jets whilst FSRQs correspond to FRII type jets (\cite{1995PASP..107..803U}). 

\subsection{Using emission spectra to understand jet dynamics and properties}

Without trustworthy black hole mass estimates for blazars it is difficult to distinguish between the two scenarios above. However, we have shown that the dynamics and geometry of a jet depends on the jet power. This is exciting because these results can be used to inform MHD simulations and allow us to better understand the physical processes responsible for jet production. For example, we can interpret the larger bulk Lorentz factor of FSRQ jets as being caused by different initial mass loading in FSRQ and BL Lac jets. In general, we would expect the final bulk Lorentz factor of a jet to depend on the initial equipartition ratio of the jet. A section of the jet plasma starts with an energy, $E=E_{M}+E_{p}$, split between magnetic and particle energies. In the simplest scenario the jet accelerates until the magnetic and particle energies are approximately equal converting magnetic energy into bulk particle energy. This gives a final bulk Lorentz factor $\gamma=E/(2E_{p})=\sigma/2$ where $\sigma$ is a measure of the initial mass loading of the jet. From this simple reasoning we might expect that jets with larger bulk Lorentz factors have a smaller initial mass loading. Our results would then suggest that higher power FSRQ jets have less mass loading than BL Lac jets.

In scenario II we expect a quantitive difference in the disc properties of FSRQs and BL Lacs. It has been suggested that at low Eddington accretion rates discs favour radiatively inefficient, advection-dominated accretion flows (ADAFs) in which a hotter thick disc surrounds the black hole (\cite{1995ApJ...452..710N}). In this case we expect the mass loading resulting from the pair production of hard X-ray disc photons and the entrainment of disc material to be more efficient for a hotter, thick ADAF. This would support the result that the bulk Lorentz factor of FSRQ jets are higher then BL Lacs because the mass loading in BL Lac jets is larger than in FSRQs. 

These results demonstrate the value of using a realistic, extended jet model of emission to relate the different spectra of blazars to the dynamical properties of jets. In the next paper in the series we will use our model to fit to a much larger sample of FSRQs and BL Lacs in order to test whether our hypothesis of a relation between jet power and radius of the transition region holds for the blazar population as whole.

\subsection{The location of the TeV emission region}

One of the most popular current questions about blazar jets is where the region responsible for the observed TeV emission is located. Most previous investigations have assumed that the emission region is compact and located at 100-1000 Schwarzschild radii from the black hole with high Doppler factors in order to explain the timescale of the TeV emission (\cite{2008ApJ...686..181F}). In our model based on radio observations of the jet of M87 we assume the high energy emission region is located at $10^{5}R_{s}$ from the black hole where the jet has a radius $2000R_{s}$. An emission region at exactly this distance is suggested by observations of a 20 day flare in 3C279 (\cite{2010Natur.463..919A}). 

\begin{figure}
	\centering
	 \includegraphics[width=8 cm, clip=true, trim=1cm 1cm 1cm 1cm]{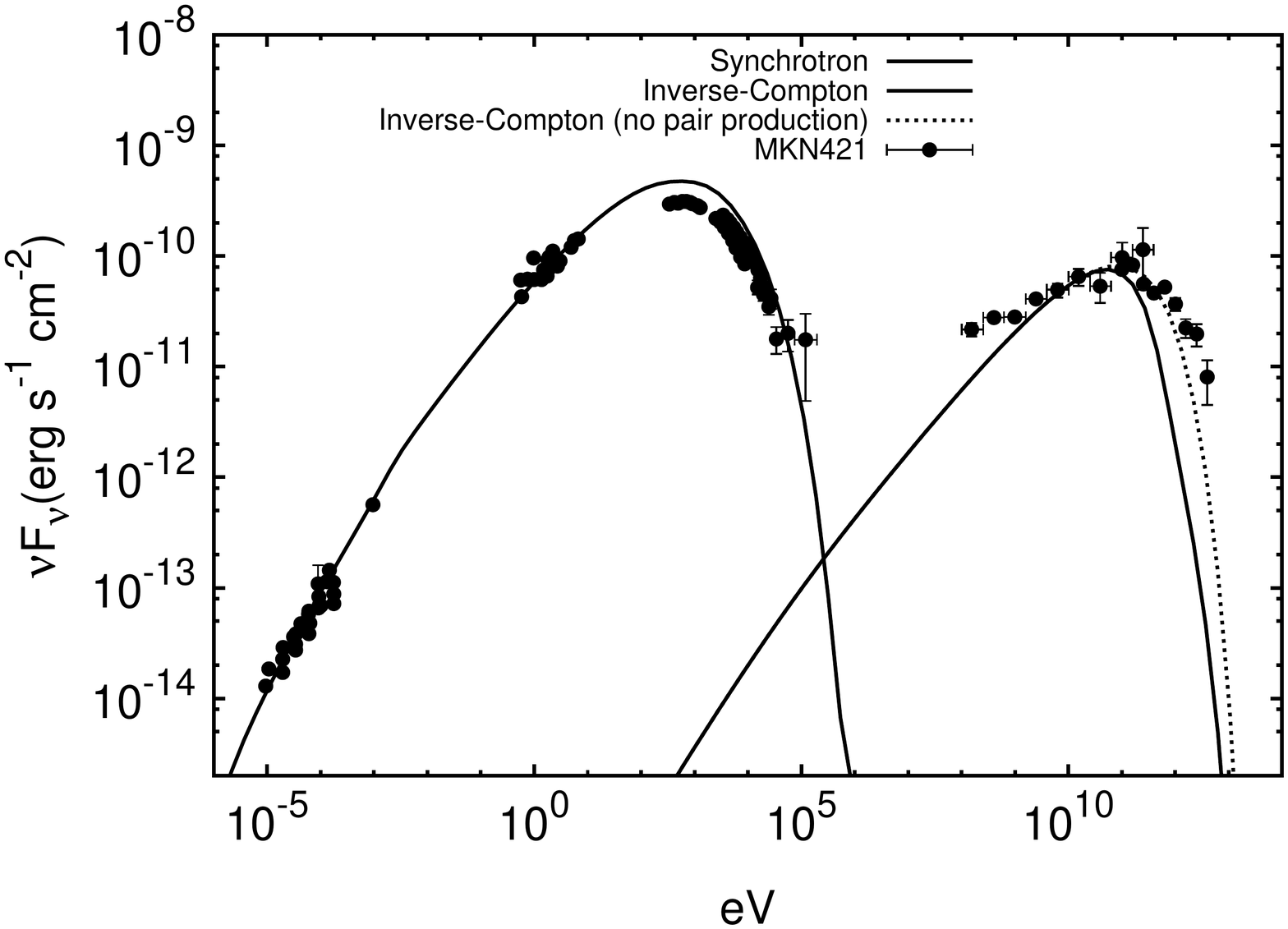} 
			
	\caption{A fit to Mkn 421 using a highly relativistic, compact, adiabatic, conical emission region, decelerating rapidly within one hundred parsecs (see Table \ref{tab2} for the parameters of the fit). We have attempted to fit the inverse-Compton emission of Mkn 421 using SSC radiation, however, we find that the large magnetic field strength required to create sufficient high energy SSC emission produces a synchrotron radiation peak at a higher energy than observed. We find that in order to produce sufficient SSC emission requires too many synchrotron photons so that the region is optically thick to TeV photons due to pair-production. }
	\label{fig:4}
\end{figure}

\begin{table}
\centering
\begin{tabular}{| c | c |}
\hline
Parameter & Mkn 421 \\ \hline 
$W_{j}$ & $3.0 \times 10^{37}\rm{W}$  \\ \hline
L & $3 \times 10^{18}\rm{m}$  \\ \hline
$E_{\m{min}}$ & 5.11 \rm{MeV} \\ \hline
$E_{\m{max}}$ & 57.3 \rm{GeV}  \\ \hline
$\alpha$ & 1.6  \\ \hline
$\theta'_{\m{opening}}$ & $3.0^{o}$ \\ \hline
$\theta_{\m{observe}}$ & $2.0^{o}$ \\ \hline
$\gamma_{\m{0}}$ & 4 \\ \hline
$\gamma_{\m{max}}$ & 50 \\ \hline
$\gamma_{\m{min}}$ & 3 \\ \hline
$R_{T}$ &  $5.9 \times 10^{13} \m{m}$\\ \hline
$M$ & $1.0 \times 10^{7}M_{\odot}$ \\ \hline
\end{tabular}
\caption{The values of the physical parameters used in the model fit shown in Figure $\ref{fig:4}$.}
\label{tab2}
\end{table}

\begin{figure*}
	\centering
		\subfloat[J0035]{ \includegraphics[width=8 cm, clip=true, trim=1cm 1cm 0cm 1cm]{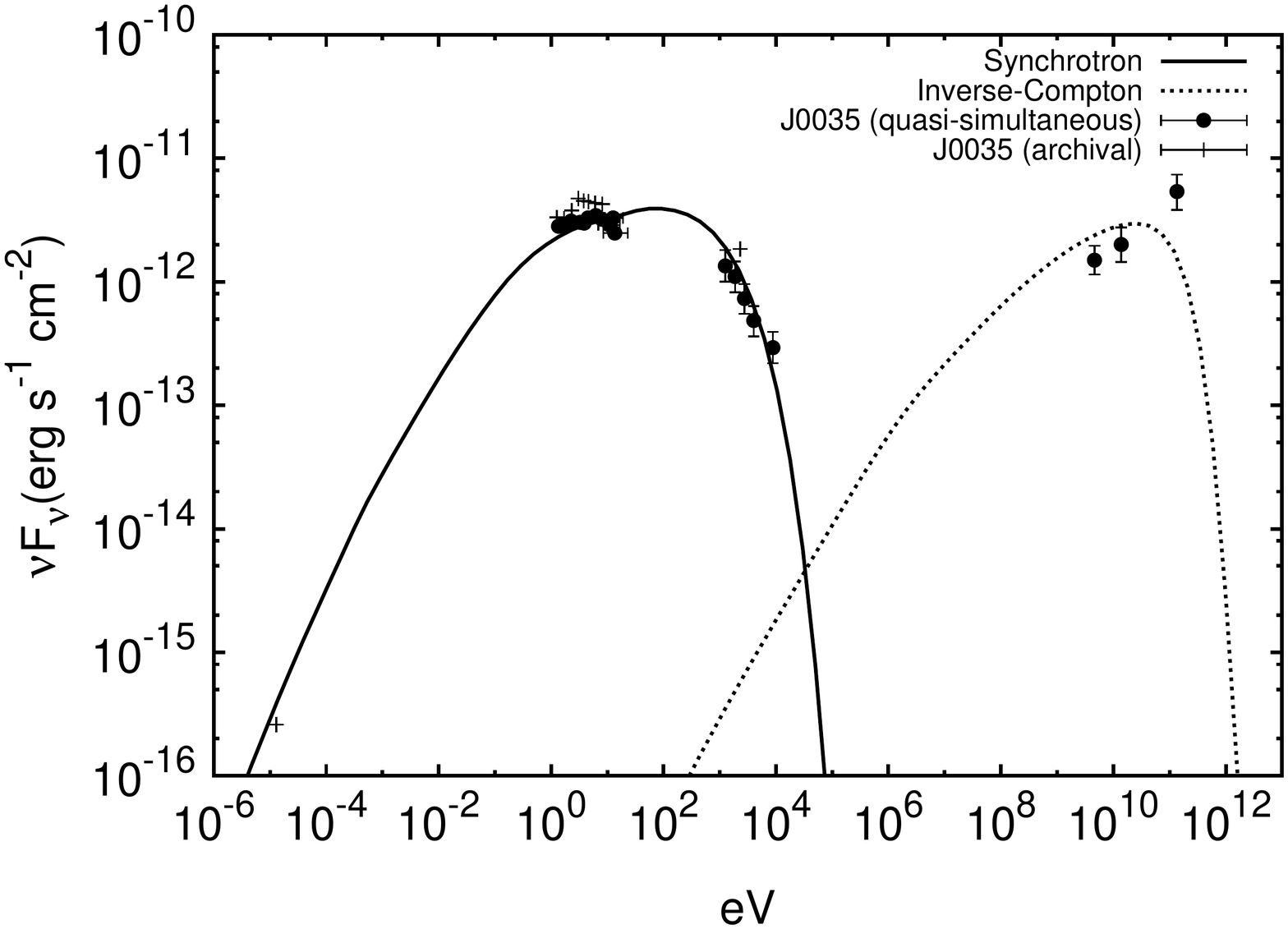} }
		\qquad
		\subfloat[J0537]{ \includegraphics[width=8cm, clip=true, trim=1cm 1cm 0cm 1cm]{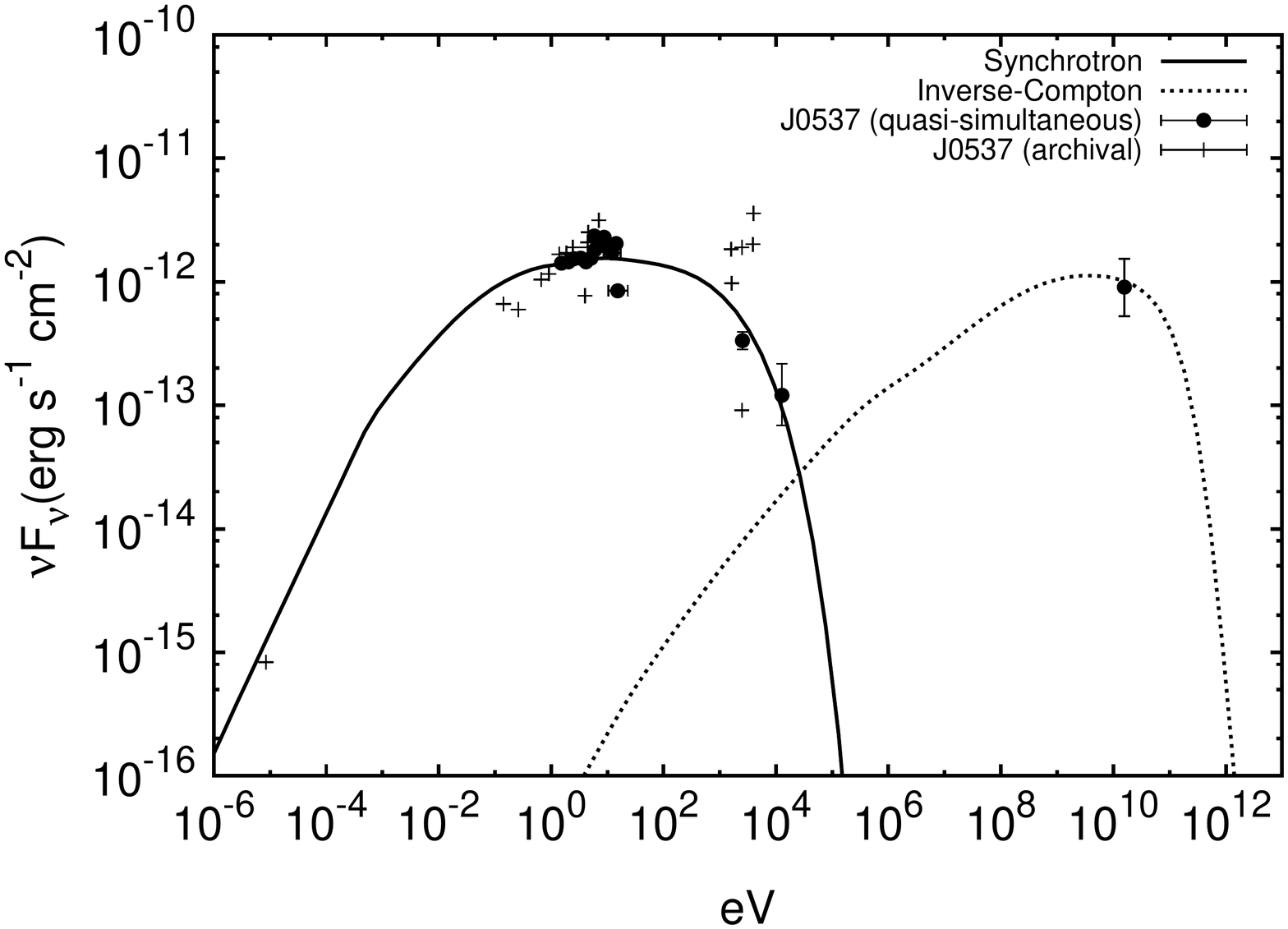} }
		\\
		\subfloat[J0630]{ \includegraphics[width=8cm, clip=true, trim=1cm 1cm 0cm 1cm]{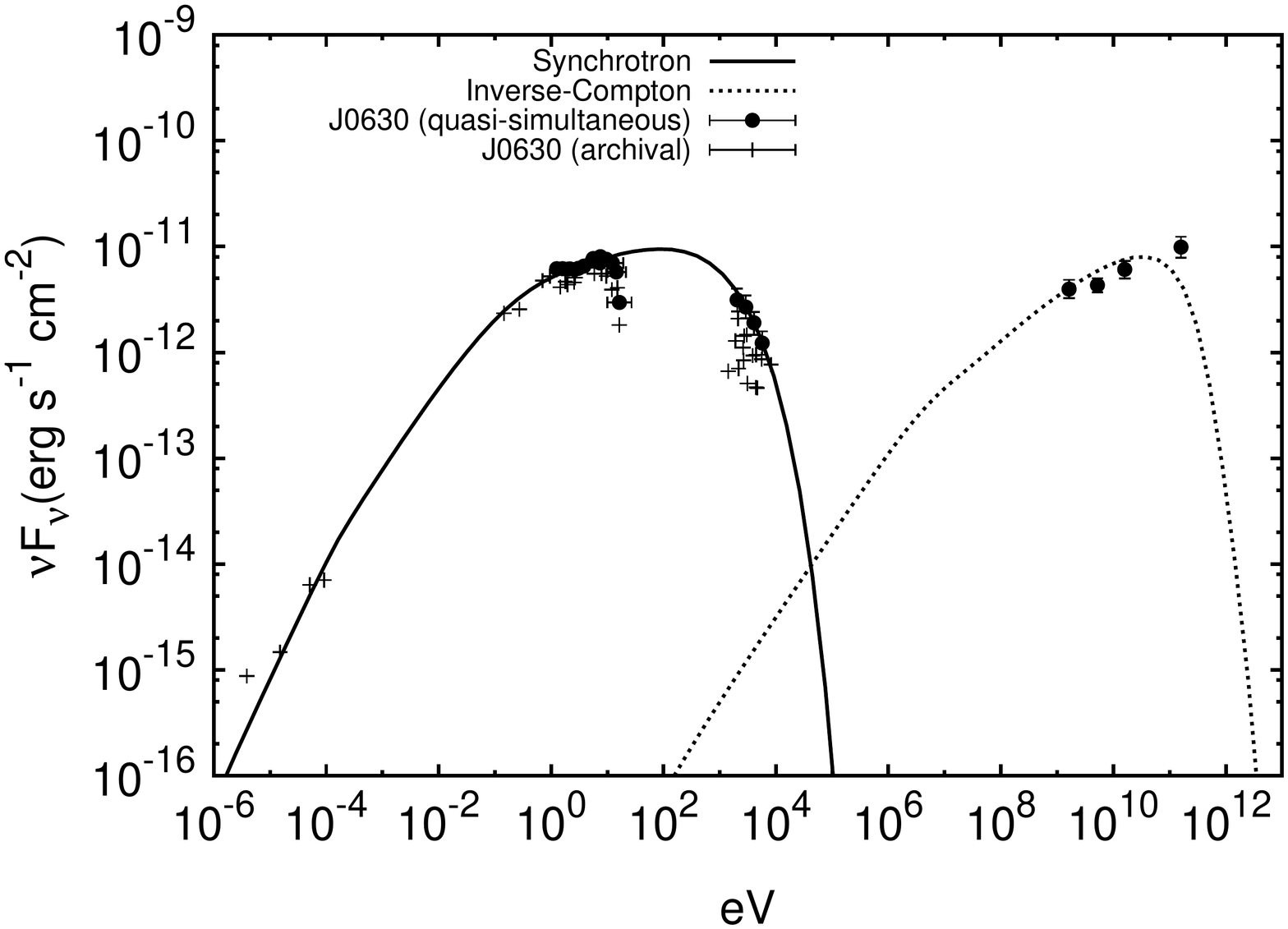} }
		\qquad
		\subfloat[J1312]{ \includegraphics[width=8 cm, clip=true, trim=1cm 1cm 0cm 1cm]{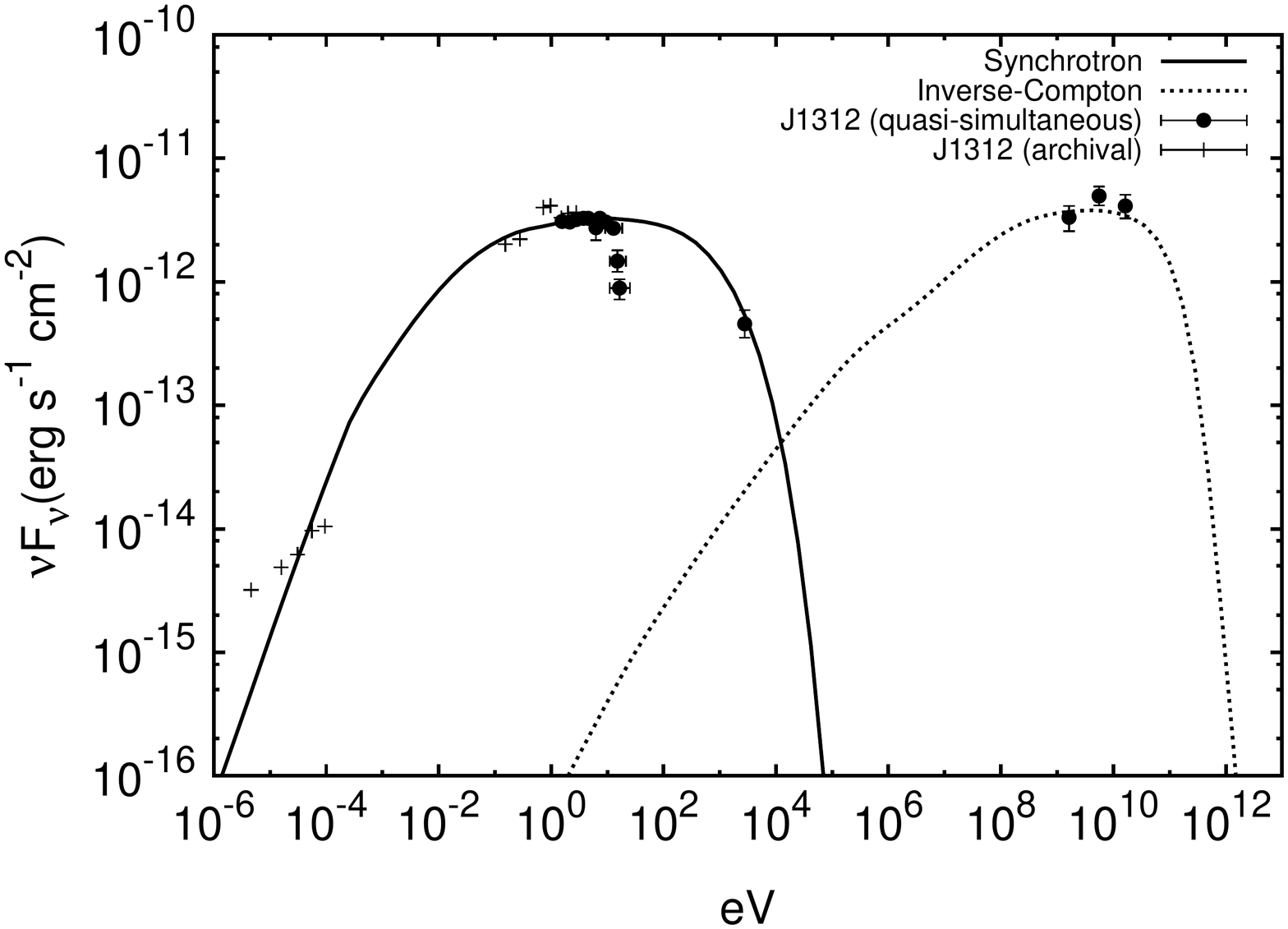} } 
			
	\caption{The results of fitting our model to the SEDs of the four high power, high synchrotron peak frequency BL Lacs J0035, J0537, J0630 and J1312 taken from Padovani et al. (2012). The data is not simultaneous and so we use these fits simply to try to understand the reason for the high peak frequencies of emission in these BL Lacs. The data shows absorption by neutral hydrogen at $\approx 10$eV which was used to estimate the photometric redshift of the blazars by Rau et al. (2012). The model fits the observations well for all blazars across all wavelengths, the parameters of the fits are shown in Table $\ref{tab3}$. We find that the high synchrotron peak frequencies are caused by high maximum electron energies whilst the remaining jet parameters are consistent with other relatively high power BL Lacs.   }
	\label{fig:5}
\end{figure*}

\begin{table*}
\centering
\begin{tabular}{| c | c | c | c | c |}
\hline
Parameter & J0035.2$+$1515 & J0537.7$-$5716 & J0630.9$-$2406 & J1312.4$-$2157 \\ \hline 
$W_{j}$ & $5.0 \times 10^{37}\rm{W}$ & $8.0\times 10^{37}\rm{W}$ & $2.1 \times 10^{38}\rm{W}$ & $2.3 \times 10^{38}\rm{W}$  \\ \hline
L & $1 \times 10^{20}\rm{m}$ & $1 \times 10^{21}\rm{m}$ & $1 \times 10^{20}\rm{m}$ & $1 \times 10^{21}\rm{m}$ \\ \hline
$E_{\m{min}}$ & 5.11 \rm{MeV} & 5.11 \rm{MeV} & 5.11 \rm{MeV} & 5.11 \rm{MeV} \\ \hline
$E_{\m{max}}$ & 40.6 \rm{GeV} & 102 \rm{GeV} & 81.0 \rm{GeV} & 81.0 \rm{GeV} \\ \hline
$\alpha$ & 1.3  & 1.75  & 1.3 & 1.6  \\ \hline
$\theta'_{\m{opening}}$ & $4^{o}$ & $4^{o}$ & $4^{o}$ & $4^{o}$ \\ \hline
$\theta_{\m{observe}}$ & $1.0^{o}$ & $1.5^{o}$ & $1.0^{o}$ & $1.2^{o}$  \\ \hline
$\gamma_{\m{0}}$ & 4 & 4 & 4 & 4  \\ \hline
$\gamma_{\m{max}}$ & 14 & 12 & 14 & 10  \\ \hline
$\gamma_{\m{min}}$ & 10 & 6 & 10 & 6  \\ \hline
$R_{T}$ & $4.6 \times 10^{14}\m{m}$ & $9.14 \times 10^{14} \m{m}$ & $2.28 \times 10^{15} \m{m}$ & $2.74 \times 10^{15} \m{m}$  \\ \hline
$M$ & $7.7 \times 10^{7}M_{\odot}$ & $1.54 \times 10^{8}M_{\odot}$ & $3.84 \times 10^{8}M_{\odot}$ & $4.62 \times 10^{8}M_{\odot}$ \\ \hline
\end{tabular}
\caption{The values of the physical parameters used in the model fits shown in Figure \ref{fig:5}.}
\label{tab3}
\end{table*}

This places the TeV emitting region at a distance 6.3pc with radius $3.8\times 10^{15}$m for Mkn421 in our model fit, approximately compatible with the upper bound on the FWHM of the radio core $<3\times 10^{15}$m of Mkn 421 found with VLBI (\cite{2006A&A...457..455C}) and at a similar distance as the flaring region $>14$pc found by \cite{2011ApJ...726L..13A} using VLBI measurements of OJ287. We find that the inverse-Compton emission cannot be due to SSC radiation because to produce a comparable amount of synchrotron and SSC emission would require a large number density of synchrotron photons, making the region optically thick to TeV photons. We find with this large emission region and small magnetic field the inverse-Compton emission is primarily due to the scattering of external CMB and starlight photons.   

Our model involves an emission region with a larger radius at a larger distance from the black hole than that generally assumed in the literature. The reason for assuming a much smaller radius is to explain the TeV flaring timescales of several minutes due to the light crossing time for the emitting region. In this paper we are only concerned with modelling the quiescent emission from blazars and we have found that this emission originates at large distances along the jet, however, our jet model is still compatible with flares occuring at shorter distances, consonant with short flaring timescales.

We find that the radio synchrotron spectrum is compatible with an optically thin single power law, implying a relatively large synchrotron bright, transition region and large inferred black hole mass $6.3\times 10^{8}M_{\odot}$ in our model. This does not preclude flares occuring at smaller distances along the jet, however, we have found that in Mkn 421 compact regions travelling at typical bulk Lorentz factors $\sim10$ are opaque to TeV photons. The common solution to this problem is to increase the Lorentz factor and radius of the compact region so that the observed Doppler-boosted light crossing time of the region is still compatible with minute timescales (e.g. bulk Lorentz factors $>50$, \cite{2008MNRAS.384L..19B}). It seems unlikely that a low power blazar like Mkn 421 with relatively slow observed superluminal motions of 2-3c (\cite{1999APh....11...19M}) should have a jet with a bulk Lorentz factor of $50$. We find that in our model fit the maximum bulk Lorentz factor is $12$.  

In Figure \ref{fig:4} we show a fit for the quiescent spectrum of Mkn 421 using a compact, adiabatic, conical jet with a high bulk Lorentz factor of 50 decelerating to a Lorentz factor of 3 at a distance of one hundred parsecs. The parameters of the fit are shown in Table \ref{tab2}. This small region with a high bulk Lorentz factor has been suggested by previous investigations and we have decelerated the jet on pc scales as suggested by \cite{2006A&A...457..455C} to explain the lack of superluminal VLBI measurements (although a lack of observed superluminal motion on pc scales does not mean a jet is not relativistic \cite{2006A&A...457..455C}). We find that even with this high bulk Lorentz factor the emission region is still optically thick to TeV photons. This is because in order for a region to emit the observed ratio of synchrotron to inverse-Compton emission through SSC it requires a large number density of synchrotron photons to scatter, however, a large number density of synchrotron photons means that the emission region is then optically thick to TeV photons. For these reasons we argue that the inverse-Compton emission observed in the quiescent spectrum of Mkn 421 is unlikely to be due to SSC radiation.

Most previous investigations have either neglected the pair production opacity due to external photons or not Lorentz transformed the photon fields into the plasma rest frame where they are Doppler-boosted. The location of the TeV emission region is difficult to resolve with current instruments, however, future instruments such as the Cherenkov Telescope Array (CTA) which provide better angular and temporal resolution, and sensitivity would be very useful in investigating the location of the TeV emission. We intend to investigate whether it is possible to put further constraints on the size and location of the TeV emitting region by modelling the flares of blazars in the near future.

\subsection{High power, high synchrotron peak BL Lacs}

Recently a number of high power BL Lacs with high synchrotron peak frequencies (HSPs) have been discovered by \cite{2012MNRAS.422L..48P}. These blazars are at odds with the general anticorrelation between the peak frequencies of emission and radio power, known as the blazar sequence. We have fitted our jet model to the four non-simultaneous spectral energy distributions (SEDs) presented in \cite{2012MNRAS.422L..48P} to understand how the properties of these blazars differ from those forming the standard blazar sequence. We show the fits to the non-simultaneous data in Figure \ref{fig:5}, the parameters of the fits are shown in Table \ref{tab3}. The observations show absorption from neutral hydrogen at $\approx 10$eV which was used by \cite{2012A&A...538A..26R} to obtain redshift estimates for the blazars. We find that the power of these BL Lacs is relatively high ($\sim 10^{38}$W) but consistent with the range of powers we find for the six BL Lacs with simultaneous multiwavelength observations. The radius of the transition region is also similar to the higher power BL Lacs in this small sample. We find that the high peak synchrotron frequencies are simply caused by the relatively high maximum energies of the non-thermal electrons ($40 - 100$GeV) but the other properties of the fits such as the jet power, bulk Lorentz factors and radius of the transition regions seem fairly typical of BL Lacs with similar jet powers. The jet properties are substantially different from the FSRQs fitted in Paper III and even from the BL Lac J1751, which appears to have many FSRQ characteristics. We find that the inverse-Compton emission of these four BL Lacs can only be fitted by SSC, due to the high peak frequency of emission and is not compatible with scattering external photons (as in FSRQs). This leads us to conclude that these blazars are, in fact, fairly typical BL Lacs, except for the relatively high maximum electron energies compared to most BL Lacs of their power. These BL Lacs have been selected from a large sample on the basis of their high peak frequency of emission and so it is not surprising that they have relatively large maximum electron energies since this quantity seems to vary considerably within the BL Lac population and strongly effects the peak synchrotron frequency ($\nu \propto \gamma_{\m{max}}^{2}B$).

In Paper III we postulated that the anticorrelation between peak frequency of emission and jet power of blazars is caused by a correlation between the radius of the transition region and the jet power. This results in higher peak frequencies of both synchrotron and inverse-Compton emission for low power blazars due to larger magnetic field strengths. We find that the transition region radius for these four high power HSP BL Lacs is larger than lower power BL Lacs and considerably smaller than the sample of high power FSRQs from Paper III. We also find that their inverse-Compton emission is produced by SSC and their bulk Lorentz factors are smaller than in the FSRQs fitted in Paper III. The properties of these BL Lacs are entirely consistent with our hypothesis.

\section{Conclusion}

In this paper we use the realistic extended jet model based on observations of the jet in M87 from Paper II to investigate six BL Lac type blazars. We calculate the absorption of high energy photons due to photon pair production by Lorentz transforming the ambient synchrotron, accretion disc, BLR, dusty torus, NLR, CMB and starlight photon distributions into the plasma rest frame and integrating the optical depth along the jet. Our model fits very well to the simultaneous multiwavelength spectra of all these blazars across all wavelengths including radio observations. We find that five out of the six BL Lac spectra are fitted by relatively low power jets with low bulk Lorentz factors and smaller radius transition regions (where the jet first comes into equipartition, stops accelerating and is brightest in synchrotron and SSC emission) compared to the powerful Compton dominant blazars fitted in Paper III. These results are consistent with our predictions from Paper III and the blazar sequence.  

We find that one of the BL Lacs, J1751, requires a large jet power, bulk Lorentz factor and radius transition region. Furthermore, we find that in order to fit both the synchrotron and high energy emission of J1751 its inverse-Compton emission must be due to scattering of low redshift CMB photons in agreement with our prediction of Compton-dominant blazars at low redshift from Paper III. We find that four out of six of the BL Lacs have gamma-ray emission best fitted by SSC.

We have tested our predictions from Paper III: blazars with large jet powers have large bulk Lorentz factors and larger radius transition regions. We find empirical evidence for this scenario using the 13 fits to blazar spectra from this work and Papers II and III. We find a clear link between the jet power of a blazar and the radius of the transition region of the jet where the plasma first comes into equipartition and becomes synchrotron bright. This leads us to an understanding of the physical basis of the blazar sequence. BL Lacs with low jet powers have larger magnetic field strengths in the synchrotron bright transition region than high power FSRQs and so have higher peak frequency synchrotron and SSC emission. FSRQs have large bulk Lorentz factors and so scatter external CMB photons at large distances leading to their Compton-dominance and lower inverse-Compton peak frequency.

We can interpret this result in one of two obvious scenarios. In the first scenario we assume jets have a universal geometry which scales linearly with black hole mass and BL Lacs have lower black hole masses than FSRQs. This scenario is sensible if most blazars accrete at a similar Eddington fraction and this leads to a prediction of a linear relation between jet power and the transition region radius. The range of black hole masses we infer from fitting this model to the sample of blazars in this and the previous paper using the observed jet geometry of M87 coincides with the range of black hole masses found from large galaxy surveys  ($10^{7}-2\times 10^{10}M_{\odot}$). In the case of FSRQs with accretion disc spectra visible above the jet emission the black hole mass found from fitting the accretion disc spectrum agrees with that inferred from our model fit if the transition region occurs at $10^{5}R_{s}$ as in M87.

If we instead assume that BL Lacs and FSRQs have similar masses (as suggested by indirect measurements of black hole mass) we arrive at a second scenario: a dichotomy of accretion modes between FSRQs and BL Lacs. In this case we find that FSRQs are accreting at a higher Eddington rate than BL Lacs as has been suggested previously. In this scenario we can intepret the different bulk Lorentz factors and length of accelerating regions as being due to different mass loading from two accretion modes. If BL Lacs contain hot, thick, radiatively inefficient discs they can more efficiently pair produce from hard X-ray photons and entrain disc material close to the jet base. This could increase the mass loading of the jet and decrease its final bulk Lorentz factor and the length required to accelerate relative to FSRQs with radiatively efficient, thin discs and lighter mass loading.

We find that in order for the quiescent high energy emission region in Mkn421 to be transparent to the observed TeV photons, we require that the inverse-Compton emission is due to scattering external photons and is not due to SSC. The radius of the transition region in our fit, $3.8 \times 10^{15}\m{m}$, implies a central black hole mass $6.3 \times 10^{8}M_{\odot}$, using a model with a jet geometry based on radio observations of the jet in M87 scaled linearly with black hole mass (\cite{2012ApJ...745L..28A}). In this case we find that the inverse-Compton emission is due to scattering of CMB photons. We also attempt to fit the spectrum of Mkn421 using a compact conical jet with a high bulk Lorentz factor, 50, decelerating rapidly over one hundred parsecs to a Lorentz factor of 3. We find that even with this high Lorentz factor SSC emission is absorbed by pair production and is not able to reproduce the observed spectrum. We find that to reproduce the correct Compton-dominance using SSC emission requires a relatively large number density of synchrotron photons to scatter, which results in the absorption of TeV photons via pair-production.

Finally, we fit the non-simultaneous multiwavelength SEDs of the four recently discovered high power, high synchrotron peak (HSP) BL Lacs from \cite{2012MNRAS.422L..48P}. We find that their high peak frequencies of emission are caused by their relatively high maximum electron energies ($40-100$GeV). The other jet properties of these blazars are typical of the higher power BL Lacs we fit to in this paper ($W_{j} \sim 10^{38}$W) and we find that their inverse-Compton emission is due to SSC. The transition region radius for these four blazars are consistent with our hypothesis of a correlation between radius and jet power. We find their bulk Lorentz factors and jet powers are significantly lower than the Compton-dominant FSRQs fitted in Paper III. We conclude that these blazars have properties typical of other high power BL Lacs and are therefore unlikely to be HSP FSRQs as has been suggested.  

This is the first time a realistic, extended jet model has been used to fit to the spectra of a sample of BL Lac type blazars. We find these results very interesting and our model provides empirical evidence for a simple monoparametric unification of blazars depending on jet power, as suggested by the blazar sequence. In the near future we are keen to investigate whether our results hold for the blazar population as a whole.

\section*{Acknowledgements}

WJP acknowledges an STFC research studentship. GC acknowledges support from STFC rolling grant ST/H002456/1. 

\bibliographystyle{mn2e}
\bibliography{Jetpaper2refs}
\bibdata{Jetpaper2refs}

\label{lastpage}

\end{document}